\documentclass{article}

\usepackage{arxiv}

\usepackage[utf8]{inputenc} 
\usepackage[T1]{fontenc}    
\usepackage{hyperref}       
\usepackage{url}            
\usepackage{booktabs}       
\usepackage{amsfonts}       
\usepackage{nicefrac}       
\usepackage{microtype}      
\usepackage{lipsum}
\usepackage{graphicx}
\graphicspath{ {./images/} }

\usepackage{url,hyperref,lineno,microtype,subcaption}
\usepackage[onehalfspacing]{setspace}
\usepackage[dvipsnames]{xcolor}
\usepackage{graphicx}
\usepackage{adjustbox} 
\usepackage{multirow}
\usepackage{tabularx}
\definecolor{mygreen}{RGB}{0,175,80}
\usepackage{booktabs}
\DeclareUnicodeCharacter{0308}{HERE!HERE!}

\title{ImageTBAD: A 3D Computed Tomography Angiography Image Dataset for Automatic Segmentation of Type-B Aortic Dissection}

\author{Zeyang Yao \\
  School of Medicine\\
  South China University of Technology\\
  Guangzhou, China 510006 \\
  \texttt{578105417@qq.com} \\
  \And
 Jiawei Zhang\\
  School of Computer Science\\
  Fudan University\\
  Shanghai, China 200433 \\
  \texttt{17110240008@fudan.edu.cn} \\
    \And
 Hailong Qiu \\
  Department of Cardiovascular Surgery\\
  Guangdong Provincial People's Hospital\\
  Guangzhou, China 510080 \\
  \texttt{hailong0518@qq.com} \\
      \And
 Tianchen Wang \\
  Department of Computer Science and Engineering\\
  University of Notre Dame \\
  IN, US 46556 \\
  \texttt{twang9@nd.edu} \\
       \And
 Yiyu Shi \\
  Department of Computer Science and Engineering\\
  University of Notre Dame \\
  IN, US 46556 \\
  \texttt{yshi4@nd.edu} \\
  \And
 Jian Zhuang \\
  Department of Cardiovascular Surgery\\
  Guangdong Provincial People's Hospital\\
  Guangzhou, China 510080 \\
  \texttt{Zhuangjian5413@163.com} \\
  \AND
  Yuhao Dong \\
  Department of Catheterization Lab\\
  Guangdong Provincial People's Hospital\\
  Guangzhou, China 510080 \\
  \texttt{barbardong1@outlook.com} \\
  \And
  Meiping Huang \\
  Department of Catheterization Lab\\
  Guangdong Provincial People's Hospital\\
  Guangzhou, China 510080 \\
  \texttt{huangmeiping@126.com} \\
      \And
  Xiaowei Xu \\
  Guangdong Cardiovascular Institute\\
  Guangdong Provincial People's Hospital \\
  Guangzhou, China 510080 \\
  \texttt{xiao.wei.xu@foxmail.com} \\
  }

\begin{document}
\maketitle
\begin{abstract}
Type-B Aortic Dissection (TBAD) is one of the most serious cardiovascular events characterized by a growing yearly incidence, and the severity of disease prognosis. 
Currently, computed tomography angiography (CTA) has been widely adopted for the diagnosis and prognosis of TBAD.
Accurate segmentation of true lumen (TL), false lumen (FL), and false lumen thrombus (FLT) in CTA are crucial for the precise quantification of anatomical features.
However, existing works only focus on only TL and FL without considering FLT.
In this paper, we propose ImageTBAD, the first 3D computed tomography angiography (CTA) image dataset of TBAD with annotation of TL, FL, and FLT.
The proposed dataset contains 100 TBAD CTA images, which is of decent size compared with existing medical imaging datasets.
As FLT can appear almost anywhere along the aorta with irregular shapes, 
segmentation of FLT presents a wide class of segmentation problems where targets exist in a variety of positions with irregular shapes.
We further propose a baseline method for automatic segmentation of TBAD.
Results show that the baseline method can achieve comparable results with existing works on aorta and TL segmentation.
However, the segmentation accuracy of FLT is only 52\%, which leaves large room for improvement and also shows the challenge of our dataset.
To facilitate further research on this challenging problem, our dataset and codes are released to the public \cite{ourdataset}.
\end{abstract}


\section{Introduction}
\begin{figure*}[!b]
\centering
\includegraphics[width=1\textwidth]{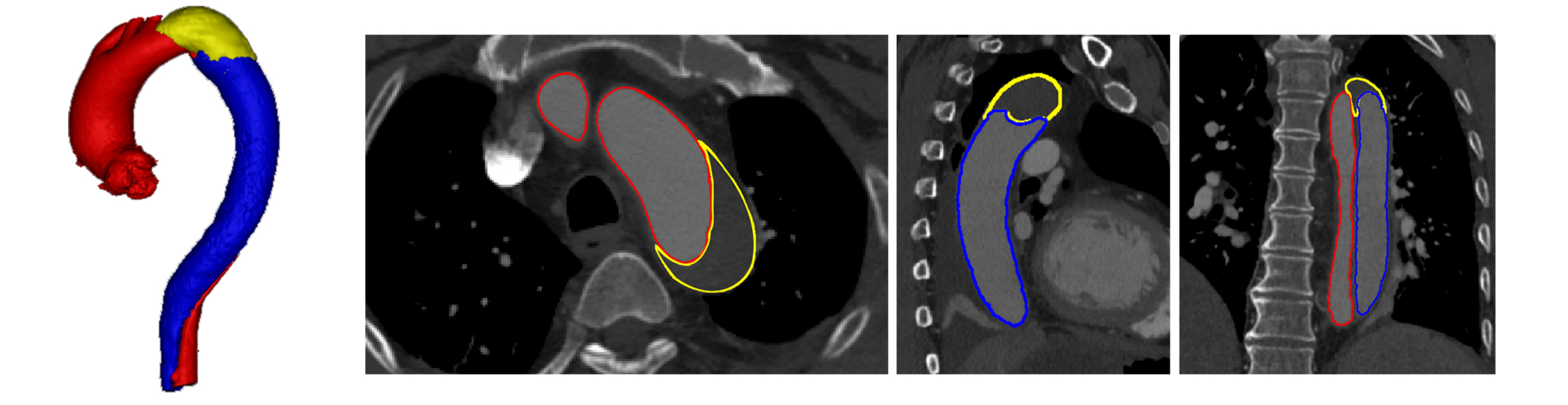}
\caption{Visualization of TBAD in a 3D model including FLT (yellow), TL (red), and FL (blue), and the corresponding CTA image with axial, coronal, and sagittal views.
}
\label{fig:display}
\end{figure*}

Type-B aortic dissection (TBAD) is the surging of blood through a tear in the aortic intima with separation of the intima and media, 
and creation of a false lumen (channel) as shown in Fig. \ref{fig:display}, which is one of the most serious cardiovascular events. 
TBAD affecting 3 to 4 per 100 000 people per year \cite{Karthikesalingam2010TheDA}. 
Approximately 20\% of patients with TBAD die before admission \cite{Karthikesalingam2010TheDA}, without treatment, 1 to 3\% patients die per hour during the first 24 hours, 30\% at the first week, 80\% at 2 weeks, and 90\% at the first year \cite{Hagan2000TheIR}. 
With the thoracic endovascular aortic repair (TEVAR) surgery and proper prognosis, patients are reportedly yielding an impressively low 30-day mortality rate of 10\% or less \cite{Hagan2000TheIR}. 
Recently, TBAD has attracted a lot of attention due to its growing yearly incidence \cite{suzuki2003clinical}, and the severity of disease prognosis.
  
Computed Tomography Angiography (CTA) is routinely adopted for the diagnosis, surgical planning, and prognosis of TBAD.
Particularly, quantification assessment of anatomical features in CTA plays a key role in surgical procedure and treatment planning for prognosis.
And segmentation of true lumen (TL), false lumen (FL) , and false lumen thrombus (FLT) is a significant step of the quantification assessment.
%
%
%
However, manual segmentation by slice is time-consuming and requires expertise, while current computer-aided approaches focus on the segmentation of the entire aorta, which is unable to segment TL, FL, or FLT, automatic segmentation of substructures of TBAD is urgently needed.
And there are already some studies trying to solve this problem. 
Specifically, Melito et al. use the adaptive algorithm together and the metamodel technique of Polynomial-Chaos Kriging
define the areas in the cross-section plane in which a point can be used to enrich the dissected segmentation for aorta dissection reconstruction.
During establishing the mathematical and computational models of aorta dissection, the level of uncertainty is extremely high. They point out that "One of the leading causes of this uncertainty is the lack of useful datasets" \cite{melito2019reliability}.  
Gamechi et al. propose a fully automatic method combining multi-atlas registration, aorta centerline extraction, and an optimal surface segmentation approach to extract the aorta surface around the centerline.
The fully automatic method they propose can assess diameters in the thoracic aorta reliably even in non-ECG-gated, non-contrast CT scans, which could be a promising tool to assess aorta dilatation in screening and in clinical practice. 
However, the method they propose still has no FLT detection ability mainly due to the lack of FLT enabled dataset \cite{gamechi2019automated}.
Particularly, there are already some works using neural networks to automatically segment TL, FL, and Aorta \cite{li2018multi, cao2019fully}.
Li et al. report a fully automatic approach based on a 3-D multi-task deep convolutional neural network that segments the entire aorta and true-false lumen from CTA images in a unified framework.
The approach they report achieves a mean dice similarity score(DSC) of 0.910, 0.849, and 0.821 for the entire aorta, true lumen, and false lumen respectively.
Cao et al. also use a convolutional neural network to solve the problems and achieves above 90\% of the mean Dice coefficients of each lumen of TBAD when not considering FLT.
They provide a promising approach for accurate and efficient segmentation of TBAD and make it possible for automated measurements of TBAD anatomical features.
However, existing works nowadays only focus on one of or both TL and FL \cite{melito2019reliability,gamechi2019automated,li2018multi,cao2019fully}, 
and FLT information is poorly explored, partially because of the lack of a dataset.
There are some other works considering thrombus in other diseases such as an abdominal aortic aneurysm (\cite{Lisowska2017ThrombusDI,Yong2017LinearregressionCN,Lopez-Linares2018}),
however, TBAD research has not yet advanced to the quantitative measurement of FLT like abdominal aortic aneurysm.

In fact, quantification assessment of FLT is also critical for surgical planning and prognosis.
First, the FLT description in clinical radiology reports plays a pivotal role in guiding the endovascular intervention surgery \cite{Dohle2017TheIO}.
Second, FLT greatly affects patients' postoperative complications \cite{Higashigaito2019AorticGA} thus is also a significant independent predictor of post-discharge mortality in prognosis \cite{Higashigaito2019AorticGA,Trimarchi2013ImportanceOF}.
Automatic, efficient, and accurate assessment of FLT is particularly useful for doctors to make a decision on TBAD.

In this paper, we propose ImageTBAD, the first 3D CTA image dataset of TBAD with annotation of TL, FL, and FLT.
For simplification of discussion, FL is the part of traditional FL without FLT in our paper.
The proposed dataset contains 100 TBAD CTA images, which is of decent size compared with existing medical imaging datasets.
Compared with TL and FL, FLT can appear in almost anywhere along the aorta with irregular shapes, which introduces many challenges to accurate segmentation of it.
FLT segmentation represents a wide class of segmentation problems where targets exist in a variety of positions with irregular shapes.
We further proposed a baseline method based on 3D U-net \cite{cciccek20163d} for automatic segmentation of TBAD.
Results show that the baseline method can achieve comparable results with existing works on the aorta and TL segmentation.
However, the segmentation accuracy of FLT is the only 52\%, which leaves large room for improvement and also shows the challenge of our dataset.
To facilitate further research on this challenging topic, our dataset and codes are released to the public \cite{ourdataset}.

\begin{table*}[!t]
\caption{Characteristics of the ImageTBAD dataset.}
\centering
\begin{tabular}{cc}
\hline
Parameters                & N                 \\ \hline
Sex = Female (\%)                & 31(31\%)          \\
Age (\textit{Mean $\pm$ SD})             & 52.5 $\pm$ 11.3       \\
Manufacturer = Philips (\%)                   & 77(77\%)          \\
Spacing between slice ($mm$) & 0.75              \\
Size of the images ($pixels$)    & $512\times 512\times$(135$-$416) \\
Typical voxel size ($mm^3$)        & 0.25$\times$0.25$\times$0.25   \\ \hline
\end{tabular}
\label{tab:parameters}
\end{table*}

\section{The Image-TBAD Dataset}\label{sec:dataset}

The ImageTBAD dataset consists of a total of 100 3D CTA images 
gathered from Guangdong Provincial Peoples' Hospital 
from January 1, 2013, to April 23, 2015. 
Images are acquired from two kinds of scanners 
(Siemens SOMATOM Force, and Philips 256-slice Brilliance iCT system), 
the characteristics of the ImageTBAD dataset is detailed in Tbale~\ref{tab:parameters}.
All the images are pre-operative TBAD CTA images whose top and bottom are around the neck and the brachiocephalic vessels, respectively, in the axial view.
The segmentation labeling is performed by a team of two cardiovascular radiologists who have extensive experience with TBAD. 
The segmentation label of each image is fulfilled by one radiologist and checked by the other. 
The time to label each image is around 1-1.5 hours. 
The segmentation includes three substructures: TL, FL, and FLT.
There are 68 images containing FLT while 32 images are free of FLT.

\begin{figure}[!htb]
\centering
\includegraphics[width=1\textwidth]{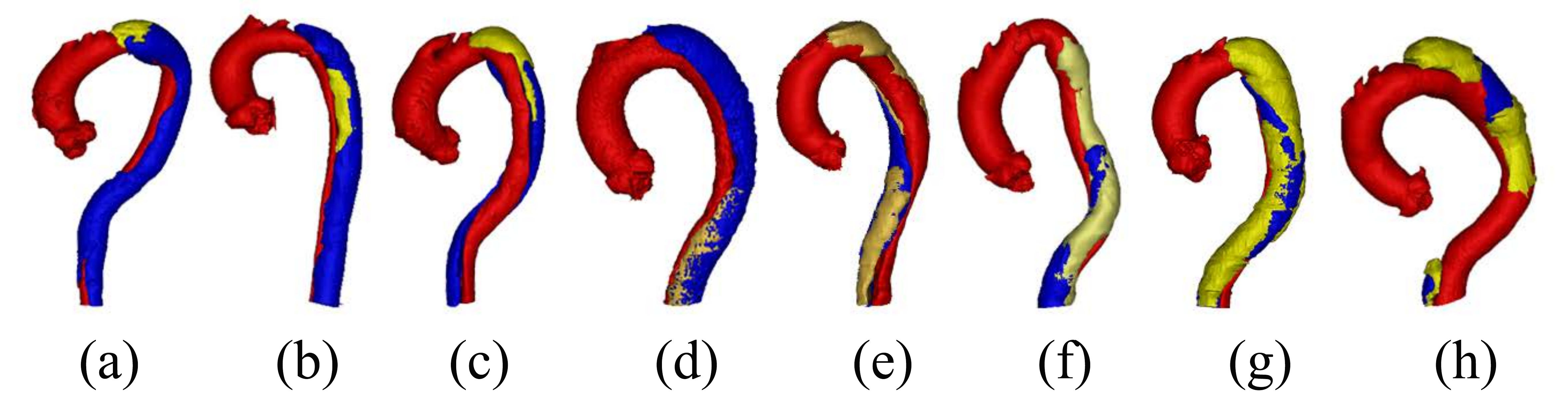}
\caption{
Examples of various relative position including (a) top, (b) middle, (c) top and middle, (d) bottom, (e-g) whole, and (h) multiple position in TBAD. The red, blue, and yellow parts correspond to TL, FL, and FLT, respectively.
Best viewed in color.
}
\label{muilti_loca_tb}
\end{figure}

By analyzing all the labels, we find the segmentation of FLT is challenging due to the following two reasons.
First, FLT can appear almost anywhere along the aorta, with irregular shapes, although most FLT appear at the top of the aorta.
Fig. \ref{muilti_loca_tb} shows a variety of relative positions of FLT. Fig. \ref{muilti_loca_tb}(a-c) shows the most common locations of FLT, while Fig. \ref{muilti_loca_tb}(d) is also common in clinic. 
Fig. \ref{muilti_loca_tb}(e-h) show some typical cases where FLT is distributed along with the whole FL and discontinued in multiple locations.
Most FLTs exist at the surface of the aorta, but there are also some located at the center of the aorta and between the FL and the TF.

\begin{figure*}[!htb]
\centering
\includegraphics[width=1\textwidth]{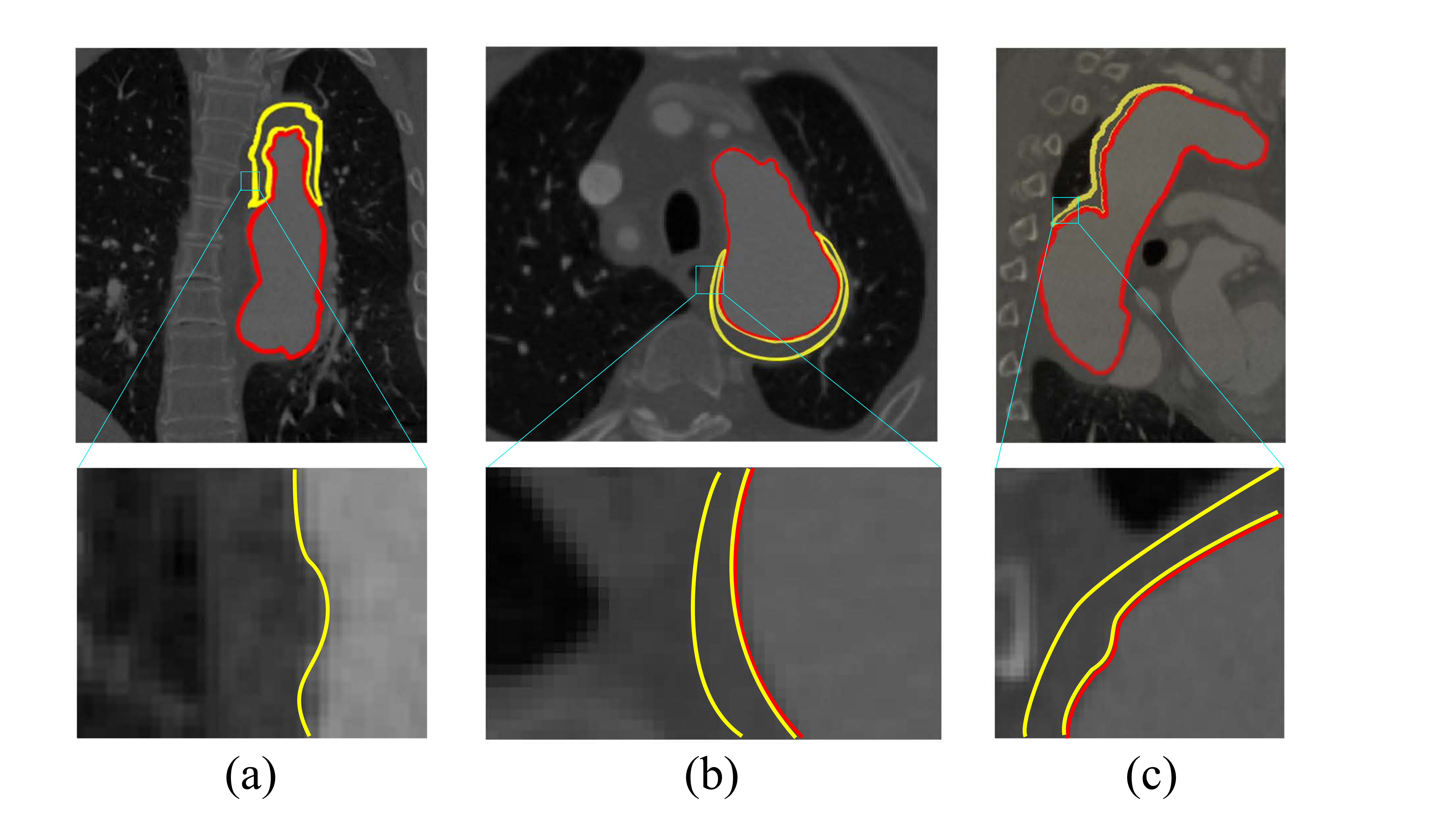}
\caption{Example of low contrast images in the ImageTBAD dataset in three views: 
(a) coronal view, (b) axial view, and (c) sagittal view.
Red and yellow lines denote to the boundary of TL and FLT, respectively.
}
\label{fig:low_contrast}
\end{figure*}

Within the eight cases in Fig. \ref{muilti_loca_tb}, we can notice the largest variety of the shapes of FLT.
Most FLTs are rather thin and long, while some others are a pile at the top of the aorta.
In addition, some FLTs are small which is relatively difficult to segment as shown in Fig. \ref{muilti_loca_tb}(g).
Second, the contrast between FLT and other tissues is rather low.
As shown in Fig. \ref{fig:low_contrast}, the intensity of the FLT and the nearby tissues are almost the same which is not easy to be visually recognized.
By zooming the area of the boundary in, we can notice some parts of the boundary as shown in Fig. \ref{fig:low_contrast}(a)(b), but some are still with high uncertainty as shown in Fig. \ref{fig:low_contrast}(c).
The low contrast would bring more challenges to FLT segmentation.

\section{Method and Experiment}
\subsection{The Baseline Method}

By analyzing the dataset, we discover the following three phenomenons.
First, the segmentation area is usually rather long in the axial view, which needs to be considered in the design of the input size.
Second, the target segmentation is rather small compared with the size of the input, and processing the whole image is not efficient.
Third, in most conditions, the combination of TL, FL, and FLT has a similar shape of the aorta. 
In fact, the part corresponding to FLT is a part of the aorta in normal anatomy.
We can also obtain FLT by removing TL and FL from the combination of the three.
This approach is expected to be more effective than direct segmentation of FLT because the complexity of shapes and positions of FLT can be avoided.
For simplicity of discussion, the combination of the three parts is donated as the aorta.

\begin{figure}[!ht]
\centering
\includegraphics[width=1\textwidth]{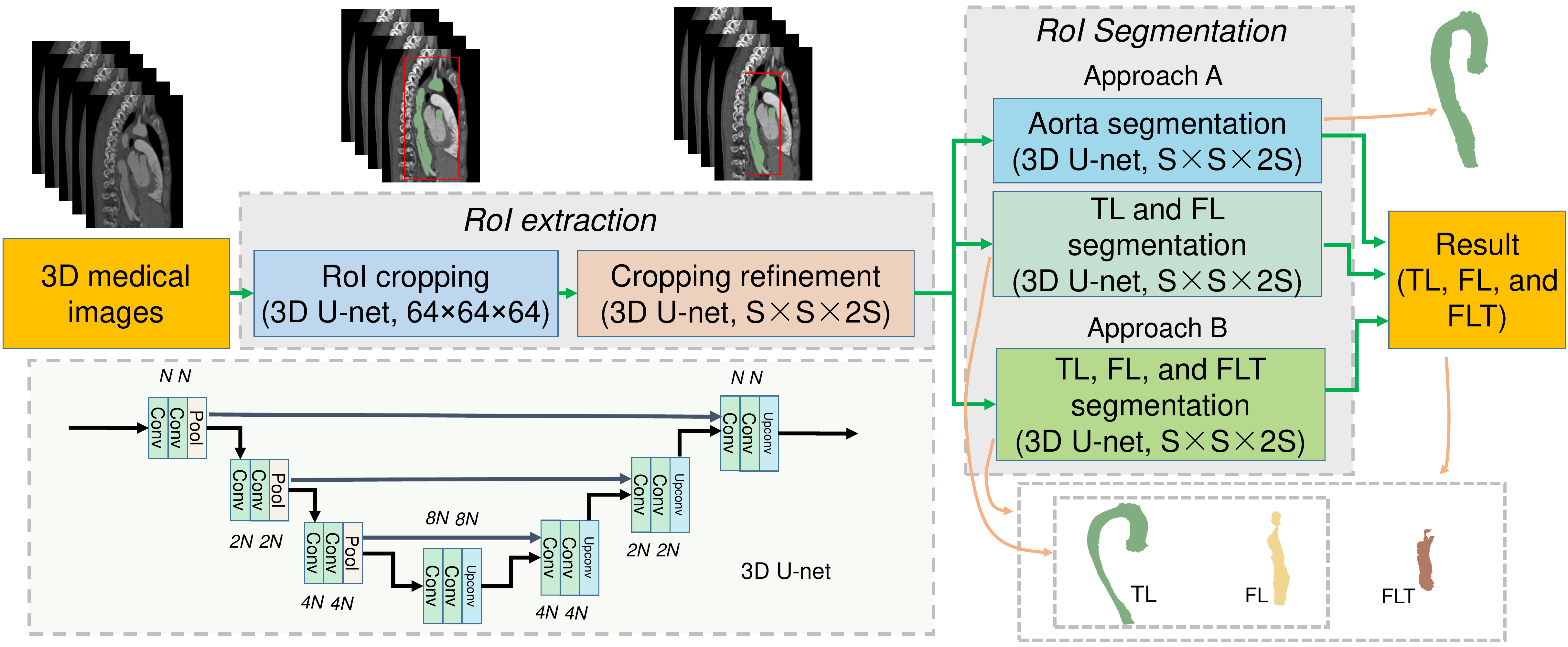}
\caption{
Processing pipeline of the proposed baseline method. 
Best viewed in color.
}
\label{fig:framework}
\end{figure}

Based on the above observations, we propose a baseline method which is a processing pipeline shown in Fig. \ref{fig:framework}.
The processing pipeline includes two steps: region of interest (RoI) extraction, and RoI segmentation.

\noindent\textbf{RoI extraction: } The RoI extraction aims to obtain a precise bounding box of the target area, which is fulfilled with two croppings.
The first cropping obtains a rough bounding box by segmenting the aorta on a resized input (original size to 64$\times$64$\times$64) using 3D U-net. 
Based on the rough bounding box, the rough RoI is cropped from the original input, and then resized to S$\times$S$\times$2S.
The cropping refinement is further proceeded on the rough RoI for aorta segmentation, and a relatively more precise bounding box of the RoI is obtained.

\noindent\textbf{RoI segmentation: } The RoI segmentation performs segmentation tasks on the refined RoI. 
We discuss two approaches: 
\textbf{Approach A}, we combine the TL and FL segmentation, and the aorta segmentation; 
\textbf{Approach B}, we perform direct segmentation of the three.
In Approach A, we suppose to easily get FLT once we obtain both TL and FL and aorta according to our discovery.
Note that all the modules adopt the same 3D U-net structure as shown in Fig. \ref{fig:framework}. 
Four resolution levels are adopted each of which contains two convolutional layers and one pooling/up-convolutional layer.
The number of filters is $N$, $2N$, $4N$, and $8N$ for the four resolution levels, respectively. 
$N$ and the input size vary for different modules as discussed above.
Post-processing only including upsampling to the original size is performed.

\subsection{Experiment}
\noindent\textbf{Experimental Setup:}
We implemented our baseline method using PyTorch based on \cite{isensee2018nnu}. 
Both Dice loss and cross entropy loss were used, and the number of training epochs was 5 for all 3D U-nets.
Data augmentation and normalization were also adopted with the same configuration as in \cite{payer2017multi} for 3D U-net.
For both networks and all the analyses, three-fold cross validation was performed (about 33 images for testing, and 67 images for training). We split the dataset so that the number of images containing FLT in each fold were the same.

We implemented two configurations, with $S=64$ and $S=96$, respectively.
Accordingly, $N=64$ and the batch size was 4 when $S=64$, and 
$N=32$ and the batch size was 3 when $S=96$.
All the experiments ran on a Nvidia GTX 1080Ti GPU with 11GB memory.

Dice score and Hausdorff distance were selected as the metrics for evaluation.
For images without FLT, the Dice score is 1 if there is no FLT in the segmentation result, otherwise 0.
As Approach B in RoI segmentation is similar to the methods that achieves the SOTA results in the TBAD \cite{li2018multi, cao2019fully}, 
we compared our method with theirs though their dataset and methods focused on the segmentation of FLT. 
Meanwhile, Hausdorff distance evaluated the shape similarity of propose method, which is formulated as follow,
$$
H(G, S) = \max \left\{ \sup _{x \in G} \inf _{y \in S}\|x-y\|, \sup _{y \in S} \inf _{x \in G}\|x-y\|\right\},
\quad \eqno{\left(1\right)}
$$
where G and S represent ground truth and prediction segmentation, respectively.\\
\textbf{Statistical Analysis:}
Differences between results are compared using the independent two-sample t-test. 
A p-value of less than 0.05 in the independent two-sample t-test is considered as statistical significance.

\begin{table}[]
\centering
\caption{
Mean and standard deviation of Dice score of baseline method, and t-test value between the Approach A and Approach B for four substructures segmentation in TBAD.}
\begin{tabular}{@{}cccccc@{}}
\toprule
        &       & Approach A & Approach B & t - value & p               \\ \midrule
S=64 & TL    & 0.82±0.08  & 0.79±0.07  & 2.327     & \textless{}0.05 \\
        & FL    & 0.72±0.20  & 0.68±0.20  & 1.166     & \textless{}0.05 \\
        & FLT   & 0.44±0.42  & 0.50±0.40  & 0.853     & \textless{}0.05 \\
        & Aorta & 0.89±0.03  &    -       &     -    &       -         \\
S=96 & TL    & 0.86±0.08  & 0.85±0.07  & 0.776     & \textless{}0.05 \\
        & FL    & 0.77±0.22  & 0.78±0.21  & 0.271     & \textless{}0.05 \\
        & FLT   & 0.44±0.43  & 0.52±0.40  & 1.123     & \textless{}0.05 \\
        & Aorta & 0.91±0.04  &    -       &     -    &       -         \\ \bottomrule
\end{tabular}
\label{table:results-dice}
\end{table}

\begin{table}[]
\centering
\caption{
Mean and standard deviation of Hausdorff distance of baseline method, and t-test value between the Approach A and Approach B for four substructures segmentation in TBAD.
}

\begin{tabular}{@{}cccccc@{}}
\toprule
        &       & Approach A & Approach B     & t - value & p          \\ \midrule
S=64    & TL    & 298.4±275.2	  & 565.0±222.4	   & 6.213     & \textless{}0.05\\
        & FL    & 597.5±1117.3   & 1089.5±1161.8	   & 2.517     & \textless{}0.05\\
        & FLT   & 1095.1±1879.9  & 1641.7±1591.8     & 1.829     & \textless{}0.05\\
        & Aorta & 300.2±273.6   &      -      & -     & -  \\
S=96    & TL    & 516.3±482.5   & 288.4±426      & 30.334     & \textless{}0.05\\
        & FL    & 1273.1±2554.4  & 643.7±1999.5	   & 1.599     & \textless{}0.05\\
        & FLT   & 1564.3±3453.5  & 978.6±2887.3     & 1.072      & \textless{}0.05\\
        & Aorta & 667.4±612.3   &     -     & -    & -  \\ \bottomrule
\end{tabular}
\label{table:results-hd}
\end{table}

\section{Results and Discussion}
\noindent\textbf{Overall results}: 

Table \ref{table:results-dice}  and Table \ref{table:results-hd} demonstrate that the mean and standard deviation of Dice score and Hausdorff distance of baseline methods (Approach A and Approach B), and their t-test value and p-value for four substructures segmentation in TBAD, respectively.
%
%
In terms of different substructures, both Approach A and Approach B achieves the highest scores on aorta with small Hausdorff distances.
However, both two methods fail to segment the TL, FL, and FLT well,  for the three are parts of the aorta without remarkable boundaries, thus relatively harder to segment them.
The Dice score and Hausdorff distance of TL beats that of FL, which may be caused by the low contrast between FL and FLT.
FLT obtains the lowest performance due to its great challenges discussed in Section \ref{sec:dataset}.
As for the two methods, though Approach A with a multi-task segmentation module achieves a bit higher Dice score with a lower Hausdorff distance than Approach B using direct segmentation, it fail to achieve higher performance on the other two parts especially on FLT.
Approach B obtains a large improvement over Approach A on FLT. This may due to the fact that direct segmentation has more constraint to more accurately define FLT than multi-task segmentation.
On the other hand, we also notice some impacts from the input size.
The Dice score of $S$=96 is slightly higher than that of $S$=64 due to the higher resolution of $S$=96.
However, the improvement is small, and there is no improvement for FLT, which indicates that higher resolution has very limited success on FLT segmentation.
Particularly, for all the 32 images without FLT, the baseline method with both configurations correctly obtain the results with Dice score of 1, which indicates that the FLT segmentation accuracy for images with FLT are much lower (about 20\%) than 52\%.

\begin{figure}[!t]
\centering
\includegraphics[width=0.99\textwidth]{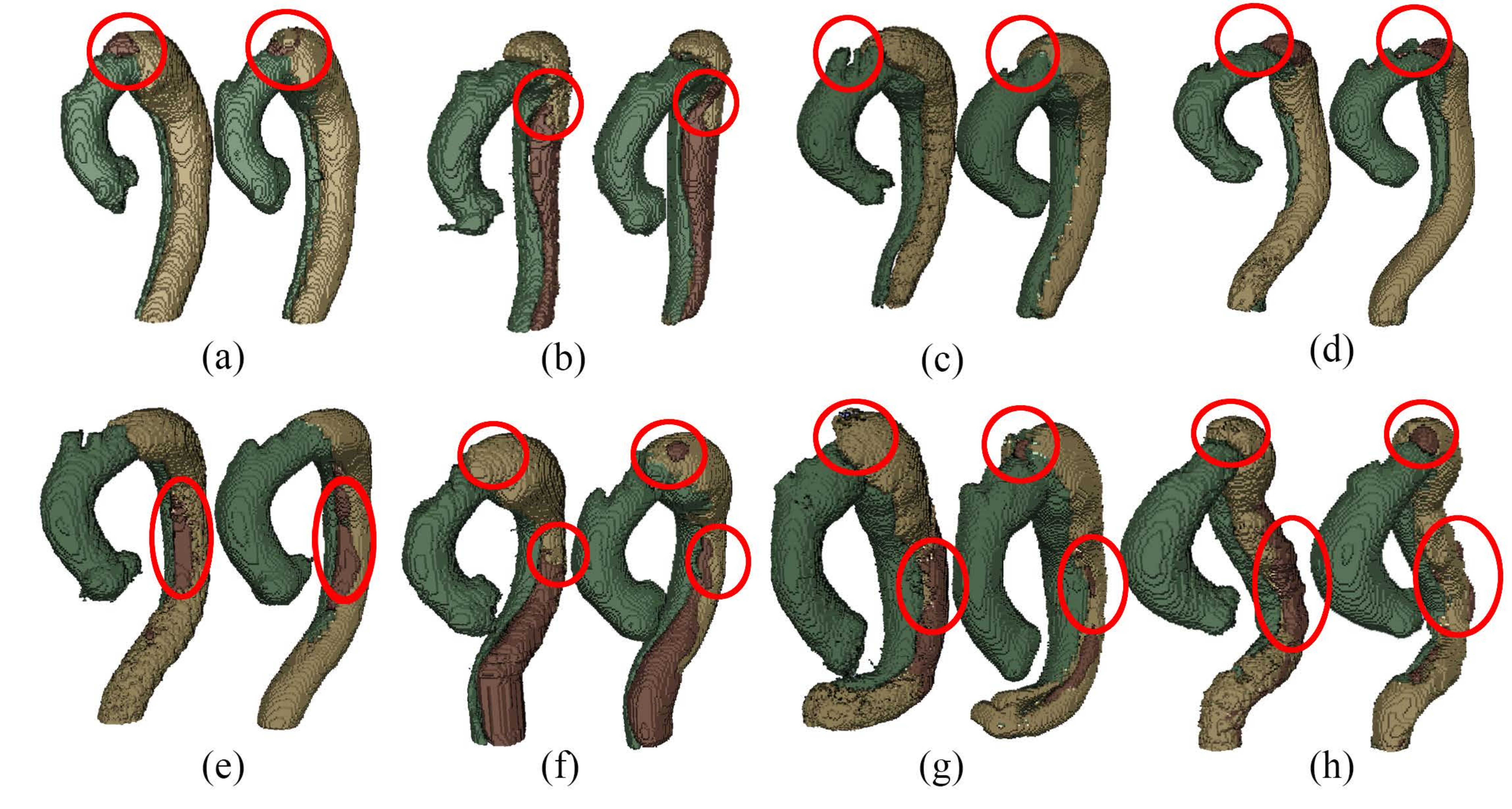}
\caption{
Examples of good segmentation results (left) with its corresponding ground truth (right).
Best viewed in color.
%
}
\label{fig:good_seg}
\end{figure}

Existing works most relevant to ours are the works proposed by a group from Tsinghua University \cite{li2018multi}\cite{cao2019fully} though the dataset and labels are different.
The method \cite{li2018multi} achieves Dice scores of 0.92, 0.85, and 0.85 on aorta, TL, and FL, respectively on the same machine (11 GB GPU memory) as ours.
The improved version \cite{cao2019fully} obtains Dice scores of 0.93, 0.93, and 0.91, on aorta, TL, and FL, respectively on a more powerful machine (32 GB GPU memory).
Compared with \cite{li2018multi}, ours achieves almost the same performance on aorta and TL, but much lower on FL.
While compared with \cite{cao2019fully}, ours obtains comparable performance only on aorta, but much lower on TL and FL. 
The comparable results on aorta indicates that our baseline method is also a powerful one.
The gaps in TL and FL may due to the difference on datasets, labels, and method details.

Though with these difference, we can still notice that accurate segmentation of FLT is rather challenging.
We hope our dataset and baseline method could help fill the gap and tackle this challenge.

\begin{figure}[!t]
\centering
\includegraphics[width=0.99\textwidth]{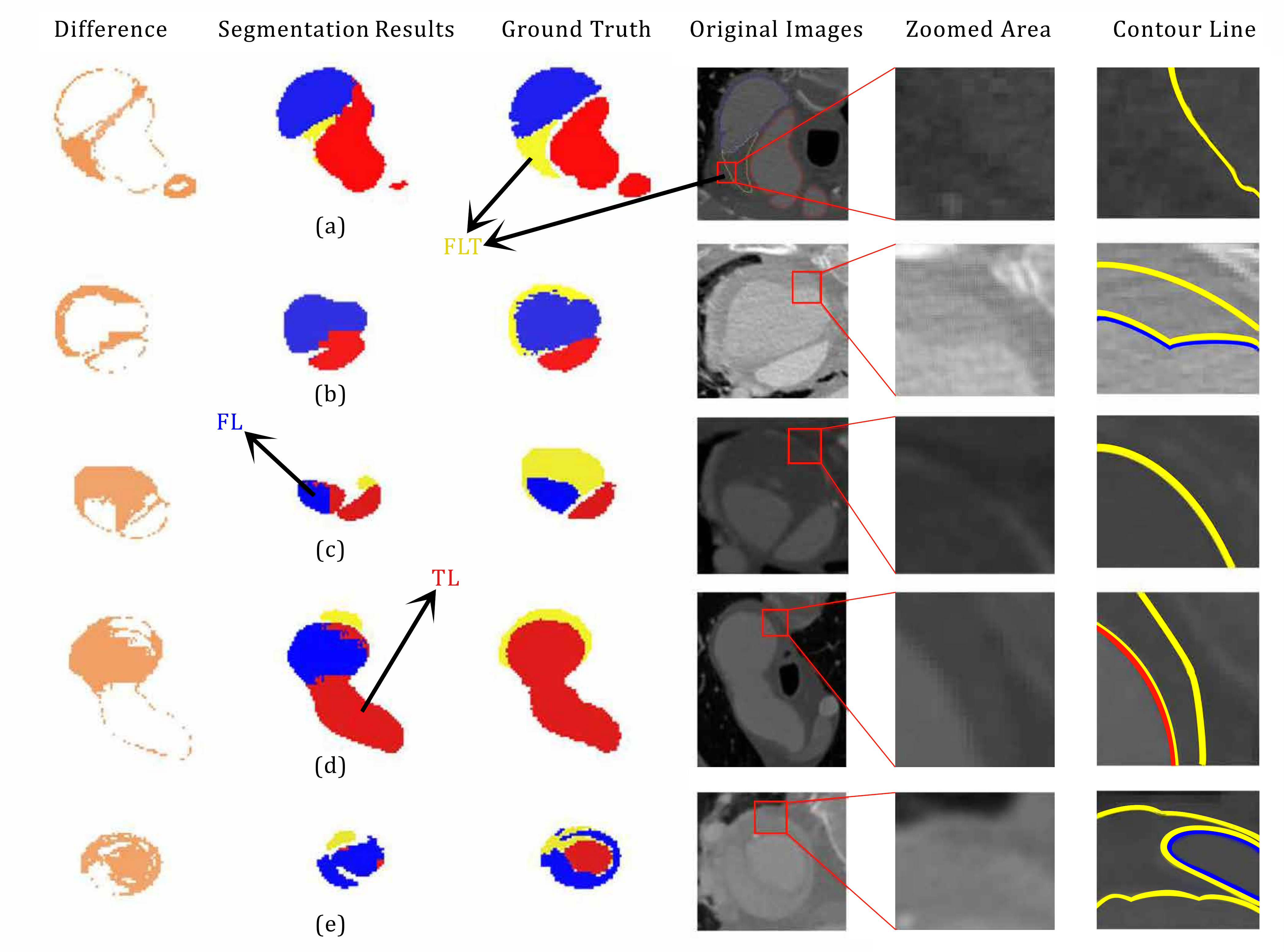}
\caption{
Examples of poor segmentation results with their corresponding ground truth, segmentation difference, original CTA image, zoomed and labeled CTA images.
The original pictures and zoomed area of each cases are accompanied.
segment failed part showed the impact of the shape and margin of thrombus in segmentation process.
Especially, case (d) is the same CT scan picture of Fig.\ref{fig:low_contrast}(b) with 180 degrees flips vertically.
Best viewed in color and position.
%
}
\label{fig:poor_seg}
\end{figure}

\noindent\textbf{Good segmentation}: 
Examples of good segmentation results are shown in Fig. \ref{fig:good_seg}.
Overall, the segmentation results have a good match with the ground truth.
However, we can still notice that compared with TL and FL, FLT has more segmentation flaws, which corresponds well to the Dice scores in Table \ref{table:results-dice}.
There is a tiny FL island at the top of the aorta which should be FLT as shown in Fig. \ref{fig:good_seg}(a).
Another three tiny FLT islands exist at the similar position which should be FL as shown in Fig. \ref{fig:good_seg}(f),  Fig. \ref{fig:good_seg}(h) and Fig. \ref{fig:good_seg}(h), respectively.
The most serious flaw of FLT is the inaccurate segmentation of its boundaries.
As shown in Fig. \ref{fig:good_seg}(b)(e)(f), there is noticeable error of the boundary segmentation.
The situations in Fig. \ref{fig:good_seg}(g)(h) is much worse, and a large part of FLT is misclassified as FL.
Most of the inaccurate boundary segmentation happens at the descending aorta.
Its low performance is usually caused by the low contrast, which also degrades the segmentation performance of FL.
TL usually has a much better performance as its contrast is much higher, and there are only some tiny flaws as shown in Fig. \ref{fig:good_seg}(c).

\noindent\textbf{Poor segmentation}:
Examples of poor segmentation results are shown in Fig. \ref{fig:poor_seg}.
Overall, there exists serious segmentation error especially for FLT.
With the context of TL and FL, the shape of FLT in Fig. \ref{fig:poor_seg}(a) can be easily recognized by humans.
However, only part of the shape is correctly segmented because of the low contrast as shown in the zoomed CTA image.
A part of FLT is lost in Fig. \ref{fig:poor_seg}(d)(e) which is due to the low contrast.
The qualities get worse in both Fig. \ref{fig:poor_seg}(b)(c) in which LFT are almost totally lost.
The boundaries is difficult to visually tell in Fig. \ref{fig:poor_seg}(b)(c).
There are also some inaccurate segmentation between TL and FL shown in \ref{fig:poor_seg}(d)(e).
The incorrect connection exists between TL and FL in Fig. \ref{fig:poor_seg}(d), and the low contrast in CTA images leads to the inaccurate segmentation between FL and TL as shown in Fig. \ref{fig:poor_seg}(e).

\section{Conclusion}
In this paper we introduce the ImageTBAD dataset to the community, which is the first 3D computed tomography angiography (CTA) image dataset of TBAD with annotation of true lumen (TL), false lumen (FL) and false lumen thrombus (FLT).
We further propose a baseline method based on 3D U-net for automatic segmentation of TBAD.
Results show that the baseline method can achieve comparable results with existing works on aorta and TL segmentations.
However, the segmentation accuracy of FLT is only 52\%, which leaves large room for improvement and proves the challenge of our dataset.
FLT segmentation represents a wide class of segmentation problems where targets exist in a variety of positions with irregular shapes.
We hope that the open-sourced code of our baseline method and dataset can encourage the community to tackle this problem.

\section*{Conflict of Interest Statement}

The authors declare no conflict of interest.

\section*{Author Contributions}

Zeyang Yao, Hailong Qiu contributed to data collection. Haiyun Yuan, Jian Zhuang, Jiawei Zhang, Qianjun Jia, Tianchen Wang, and Yiyu Shi contributed to analysis and writing. Meiping Huang, Yuhao Dong, and Xiaowei Xu
contributed to project planning, development, discussion and writing.

\section*{Funding}
This work was supported by the National key Research and Development Program of China (No. 2018YFC1002600), the Science and Technology Planning Project of Guangdong Province, China (No. 2017B090904034, No. 2017B030314109, No. 2018B090944002, No. 2019B020230003), Guangdong Peak Project (No. DFJH201802), the National Natural Science Foundation of China (No. 62006050).

\section*{Ethical approval}
This work was approved by the Research Ethics Committee of Guangdong General Hospital, Guangdong Academy of Medical Science under Protocol No. 20140316.
All procedures performed in studies involving human participants were in accordance with the ethical standards of the institutional and/or national research committee and with the 1964 Declaration of Helsinki and its later amendments or comparable ethical standards.

\bibliographystyle{unsrt}  
\bibliography{references.bib}  






\end{document}